\documentclass[a4paper,11pt]{article}
\usepackage[english]{babel}
\usepackage{graphicx}
\usepackage{mathtext}
\usepackage{indentfirst}
\usepackage{epsfig,amsmath,amsfonts}
\usepackage{amsmath}
\usepackage{amsfonts}
\usepackage{comment}
\usepackage{verbatim}
\usepackage[normalem]{ulem}
\usepackage{cancel}

\usepackage{cmap}                   % поиск в PDF
\usepackage[T2A]{fontenc}           % кодировка
\usepackage[utf8]{inputenc}         % кодировка исходного текста
\usepackage[left=2cm,right=2cm,top=1cm,bottom=2cm,bindingoffset=0cm]{geometry}
\usepackage[usenames]{color}
\usepackage{colortbl}
\usepackage[dvipsnames]{xcolor}
\usepackage{lipsum}
\usepackage{tcolorbox}
\usepackage{amsmath,amssymb}
\usepackage{blindtext}
\usepackage{titlesec}
\usepackage{hyperref}
\usepackage{cancel}
\titleformat{\section}{\normalfont\large\bfseries}{\thesection}{1em}{\,}
\titleformat{\subsection}{\normalfont\normalsize\bfseries}{\thesubsection}{1em}{\,}

\usepackage{mathrsfs}

\newcommand{\RNumb}[1]{\uppercase\expandafter{\romannumeral #1\relax}}

\usepackage{tcolorbox}
\definecolor{mycolor}{rgb}{0.122, 0.435, 0.698}

\newtcbox{\mybox}{nobeforeafter,colframe=mycolor,colback=mycolor!10!white,boxrule=0.5pt,arc=4pt,
    boxsep=0pt,left=6pt,right=6pt,top=6pt,bottom=6pt,tcbox raise base}

\PassOptionsToPackage{unicode}{hyperref}
\PassOptionsToPackage{naturalnames}{hyperref}

\usepackage[active]{srcltx}
\usepackage{tikz}
\usepackage{tikz-cd}
\usetikzlibrary{babel}
\usetikzlibrary{cd}

\newcommand{\be}{\begin{equation}}
    \newcommand{\ee}{\end{equation}}
\newcommand{\bea}{\begin{eqnarray}}
    \newcommand{\eea}{\end{eqnarray}}

\usepackage{graphicx}

\title{Unconstrained ${\cal N}=2$ higher-spin gauge
    superfields and their hypermultiplet couplings}
\date{\today}
\begin{document}
%   \begin{flushright}
%       \today
%   \end{flushright}

%\maketitle

\begin{center}

    \hfill  {}

    %\vskip 1.5cm

    {\Large \bf  Unconstrained ${\cal N}=2$ higher-spin gauge
        superfields
        \vspace{0.2cm}
        and their hypermultiplet couplings}
    \vspace{0.3cm}

    %\vspace{1cm}
    \renewcommand{\thefootnote}{$\star$}

    {\large\bf Ioseph Buchbinder} ${}^{\dag, \, \ast}$, {\quad \large\bf
        Evgeny~Ivanov} ${}^{\ast,\, \star}$, {\quad \large\bf
        Nikita~Zaigraev} ${}^{\ast,\, \star}$
%   \vspace{1.3cm}

     \vspace{0.5cm}

    {${}^\dag$ \it Center of Theoretical Physics, Tomsk State Pedagogical University,} \\
    {\it 634061, Tomsk,  Russia} \\

    \vskip 0.15cm

%   { ${}^+$ \it National Research Tomsk State University, 634050, Tomsk, Russia} \\
%   \vskip 0.15cm

    {${}^\ast$ \it Bogoliubov Laboratory of Theoretical Physics, JINR,}\\
    {\it 141980 Dubna, Moscow region, Russia} \\
    \vskip 0.15cm

    {${}^\star$ \it Moscow Institute of Physics and Technology,}\\
    {\it 141700 Dolgoprudny, Moscow region, Russia}
    \vspace{0.2cm}

    {\tt joseph@tspu.edu.ru, eivanov@theor.jinr.ru, nikita.zaigraev@phystech.edu}\\

    \vspace{0.4cm}
    
    \textit{Talk given by N. Zaigraev at  the 
    	\\
    	International Conference on Quantum Field Theory, High-Energy Physics, and Cosmology,
    \\
    Dubna, July 18 - 21, 2022}
    
    \medskip

    \begin{abstract}
    %   \noindent
        In recent papers \cite{Buchbinder:2021ite, Buchbinder:2022kzl}, we constructed free  off-shell $\mathcal{N}=2$ supersymmetric higher spin gauge theories in harmonic superspace and their cubic couplings to hypermultiplet. The present report is a brief review of the results obtained.
    \end{abstract}
    \vskip 1cm

\end{center}
%\tableofcontents

%\newpage

\section{Preamble: supersymmetry and higher spins}

             Supersymmetric higher-spin theories are under intensive development for last decades. One of the basic origins of interest in them is that they could provide
            a bridge between superstring theory and low-energy (super)gauge theories.

            Free massless bosonic and fermionic higher spin field theories have
            been pioneered in \cite{FronsdalInteg, FronsdalHalfint}.
         The natural tools to deal with supersymmetric theories are off-shell superfield methods.
            In the superfield approach the supersymmetry is closed on the off-shell supermultiplets
            with the correct sets of the auxiliary fields and so is manifest. Unconstrained superfield formulations are most preferable.
             The component approach to the description of $4D$, ${\cal N}=1$ supersymmetric
            free massless higher spin models was initiated in \cite{Courtright, Vasiliev}.
         The complete off-shell Lagrangian formulation of $4D$
            free higher spin ${\cal N}=1$ models (including those on the AdS background) has been given in terms of
            ${\cal N}=1$ superfields in a series of works by S. Kuzenko with collaborators \cite{Kuz1, Kuz2, Kuz3}.

     Until present, an off-shell superfield Lagrangian formulation for higher-spin {\bf extended}
    supersymmetric theories, with all supersymmetries manifest,  was unknown
    even for free theories.
    This gap was recently filled in \cite{Buchbinder:2021ite}.
    An off-shell manifestly ${\cal N}=2$
    supersymmetric unconstrained formulation of $4D, {\cal N}=2$ superextension of the Fronsdal theory for integer spins was constructed for the first time,
    based on the harmonic superspace approach \cite{18, HSS}.
 Manifestly ${\cal N}=2$ supersymmetric off-shell cubic couplings of $4D, {\cal N}=2$ to the hypermultiplets
    were given in \cite{Buchbinder:2022kzl}.
     Our papers open a new area of applications of the harmonic superspace formalism, that time in ${\cal N}=2$ higher-spin theories.

\section{Harmonic superspace}

            At present, in four-dimensions, the self-consistent off-shell superfield formalism for ${\cal N}=2$
            and ${\cal N}=3$ theories
            %in {\color{blue}$4D$}.
            is the harmonic superspace approach \cite{18, HSS}.
         Harmonic ${\cal N}=2$ superspace  is obtained by adding $SU(2)$ harmonics to ordinary superspace:
        \bea \label{HSS}
                Z = (x^m\,,\;\theta^\alpha_i\,, \;\bar\theta^{\dot\alpha\,j}, u^{\pm i}), \quad u^{\pm i} \in SU(2)/U(1),  \;\quad u^{+ i}u^-_i = 1.
                \eea

         Analytic harmonic ${\cal N}=2$ superspace is an invariant subspace of \eqref{HSS}:
            \bea
                &&\zeta_A = (x^m_A, \theta^{+ \alpha}, \bar\theta^{+ \dot\alpha}, u^{\pm i}), \;\quad
             \theta^{+ \alpha, \dot\alpha} := \theta^{\alpha, \dot\alpha i} u^+_i,
                \;\quad x^m_A := x^m - 2i\theta^{(i}\sigma^m \bar\theta^{j)}u^+_iu^-_j\,.
                %\nonumber
                \eea All basic ${\cal N}=2$ superfields are analytic:
            \bea
                \underline{\rm matter \;hypermultiplets}:&& q^{+}(\zeta_A)\,, \,\tilde{q}^{+}(\zeta_A)\,, \nonumber \\
                \underline{\rm SYM}:&& V^{++}(\zeta_A)\,, \;\;   \nonumber \\
                \underline{\rm supergravity}:&& h^{++ m}(\zeta_A)\,,\,h^{++ \alpha +}(\zeta_A)\,, \,h^{++ \dot{\alpha} +}(\zeta_A)\,, \,h^{++ 5}(\zeta_A)\,.  \nonumber
                \eea

\section{${\cal N}=2$ spin 1 multiplet}

            An instructive example is supplied by Abelian ${\cal N}=2$ gauge theory,\bea
                V^{++}(\zeta_A)\,, \quad \delta_\lambda V^{++} = D^{++}\lambda (\zeta_A)\,, \;\quad D^{++} = \partial^{++} - 2i\theta^{+\alpha}\bar\theta^{+\dot\alpha}\partial_{\alpha\dot\alpha} + \theta^{+\hat{\rho}}\partial^+_{\hat{\rho}}\,. %\nonumber
                \eea

             In the Wess-Zumino gauge the analytic gauge superfield $V^{++}(\zeta_A)$ has the expansion:
             \bea
                &&
                V^{++}(\zeta_A) = (\theta^+)^2 \phi + (\bar\theta^+)^2 \bar\phi + 2i\theta^{+\alpha}\bar\theta^{+\dot\alpha} A_{\alpha\dot\alpha} \nonumber \\
                && +\, (\bar\theta^+)^2 \theta^{+\alpha}\psi_\alpha^i u^-_i +
                (\theta^+)^2 \bar\theta^{+}_{\dot\alpha}\bar\psi^{\dot\alpha i} u^-_i +(\theta^+)^2(\bar\theta^+)^2 D^{(ik)}u^-_iu^-_k \,.%\nonumber
                \eea
            $4D$ fields $\phi,\, \bar\phi\,, \,A_{\alpha\dot\alpha}\,, \,\psi_\alpha^i\,, \, \bar\psi_{\dot\alpha}^i \,, \,D^{(ik)}$
            constitute an Abelian gauge ${\cal N}=2$ off-shell multiplet ( 8 + 8 off-shell degrees of freedom).

             Gauge-invariant action is written as
             \bea
                 S_{(s=1)} \sim \int d^{12}Z\,\, \big(V^{++} V^{--}\big)\,, %\nonumber %\\
             \eea
             where $V^{--}$ is a solution of zero curvature equation:
             \bea
                 D^{++} V^{--} - D^{--} V^{++} = 0\,, \;\quad \delta_\lambda V^{--} = D^{--}\lambda\,. \nonumber
                \eea
                \begin{equation*}
                    D^{--} = \partial^{--} - 2i\theta^{-\alpha}\bar\theta^{-\dot\alpha}\partial_{\alpha\dot\alpha} + \theta^{-\hat{\rho}}\partial^-_{\hat{\rho}}\,.
            \end{equation*}

\section{${\cal N}=2$ spin 2 multiplet: linearized ${\cal N}=2$ Einstein supergravity}

         In the linearized $\mathcal{N}=2$ supergravity, the analogs of gauge prepotential $V^{++}(\zeta_A)$ are the following set of analytic gauge prepotentials:
            \bea
                && \Big( h^{++m}(\zeta_A)\,, \;h^{++5}(\zeta_A)\,, \; h^{++\hat{\mu}+}(\zeta_A) \Big),  \quad \hat{\mu} = (\mu\,, \dot{\mu})\,,  \\
                &&\delta_\lambda h^{++m } = {D}^{++} \lambda^m + 2i \big( \lambda^{+\alpha} \sigma^m_{\alpha\dot{\alpha}} \bar{\theta}^{+\dot{\alpha}}
                + \theta^{+\alpha} \sigma^m_{\alpha\dot{\alpha}} \bar{\lambda}^{+\dot{\alpha}}\big)\,, \nonumber \\
                &&\delta_\lambda h^{++5} = {D}^{++} \lambda^5 - 2i \big(\lambda^{+{\alpha}} \theta^{+}_{\alpha} - \bar\theta^{+}_{\dot{\alpha}}\bar\lambda^{+\dot{\alpha}}\big)\,, \nonumber \\
                &&\delta_\lambda h^{++\hat{\mu}+} = {D}^{++} \lambda^{+\hat{\mu}}\,.  \nonumber
                \eea
              The off-shell content of this theory is revealed in the Wess-Zumino gauge::
        \begin{eqnarray}
                    &&h^{++m}
                    =
                    -2i \theta^+\sigma^a \bar{\theta}^+ \Phi^m_a
                    +  \big[(\bar{\theta}^+)^2 \theta^+ \psi^{m\,i}u^-_i + c.c.\big]+ \ldots  \nonumber \\
                    &&h^{++5} =
                    -2 i \theta^+ \sigma^a \bar{\theta}^+ C_a + \ldots\,, \quad h^{++\mu+} = \ldots  \nonumber
            \end{eqnarray}
            %where $\dots $  stand for auxiliary fields.
             The residual gauge freedom is given by:
            \begin{eqnarray}
                    \lambda^m \;\Rightarrow\; a^m(x)\,,\quad \; \lambda^5 \, \Rightarrow \; b(x)\,, \quad\;
                    \lambda^{\mu+} \;\Rightarrow \; \epsilon^{\mu i}(x) u^+_i + \theta^{+\nu}l_{(\nu}^{\;\;\;\mu)}(x)\,.\nonumber
                    %&& \lambda^{\mu+} \;\Rightarrow \; \epsilon^{\mu i}(x) u^+_i + \theta^{+\nu}l_{(\nu}^{\;\;\;\mu)}(x)\,, \,
                    %        \bar{\lambda}^{\dot{\mu}+} \;\Rightarrow \; \bar{\epsilon}^{\dot{\mu} i}(x) u^+_i + \bar{\theta}^{+\dot{\nu}}l_{(\dot{\nu}}^{\;\;\;\dot{\mu})}(x)\,. \nonumber
            \end{eqnarray}

            %One ends with the standard field content {\color{blue}$40 + 40$} of minimal {\color{blue}${\cal N}=2$} Einstein  supergravity.
            The physical fields are $\Phi^m_a, \psi^{m\,i}_\mu, C_a$ \big($\mathcal{N}=2$ multiplet $ (2, 3/2, 3/2, 1)$ on shell\big).
            %the remaining ones are auxiliary.
            The spin 1 part of $\Phi^m_a$
            can be gauged away by the local ``Lorentz'' parameters $l_{(\nu\mu)}^{\;\;\;}(x),\;   l_{(\dot{\nu}\dot{\mu})}^{\;\;\;}(x)$:
            %In this gauge
        \bea
                && \Phi^m_a \sim \Phi_{\beta\dot\beta\alpha\dot\alpha} \Rightarrow \Phi_{(\beta\alpha)(\dot\beta\dot\alpha)}
                + \varepsilon_{\alpha\beta}\varepsilon_{\dot\alpha\dot\beta} \Phi\,.\nonumber
                %&& \delta \Phi_{\beta \dot{\beta}\alpha \dot{\alpha} } =\frac{1}{2} \left(\partial_{\alpha\dot{\alpha}} a_{\beta\dot{\beta}}
                \eea

 The invariant action is constructed as \cite{Zupnik:1998td}:
            \bea
                 S_{(s=2)} \sim \int d^4x d^8\theta du \Big( G^{++\alpha\dot\alpha}G^{--}_{\alpha\dot\alpha} + G^{++5}G^{--5}\Big)\,.
            \eea

    Here $G$ superfields are composite non-analytic objects:
            \bea
                &&G^{++ \mu\dot\mu} := h^{++ \mu\dot\mu} + 2i\big( h^{++\mu+}\bar\theta^{-\dot\mu} + \theta^{-\mu}h^{++\dot\mu+}\big)\,, \nonumber \\
                &&G^{++ 5} := h^{++ 5} - 2i\big( h^{++\mu+}\theta^-_\mu - \bar\theta^-_{\dot\mu}h^{++\dot\mu+}\big)\,,\nonumber \\
                && D^{++}G^{--\mu\dot\mu} = D^{--}G^{++\mu\dot\mu}\,,\nonumber \\ && D^{++}G^{--5} = D^{--}G^{++5}\,. \nonumber
                \eea
         After passing to components, the spin 2 part of the Lagrangian reads:
            \bea
                &&G^{++\alpha\dot\alpha}_{(\Phi)}G_{(\Phi)}^{--\alpha\dot\alpha} +  G_{(\Phi)}^{++5}G_{(\Phi)}^{--5} \quad \Rightarrow \nonumber\\
                && {\cal L}_{(\Phi)} =
                -\frac{1}{4}\Big[\Phi^{(\alpha\beta)(\dot{\alpha}\dot{\beta})} \Box
                \Phi_{(\alpha\beta)(\dot{\alpha}\dot{\beta})} -
                \Phi^{(\alpha\beta)(\dot{\alpha}\dot{\beta})}
                \partial_{\alpha\dot{\alpha}} \partial^{\rho\dot{\rho}}
                \Phi_{(\rho\beta)(\dot{\rho}\dot{\beta})} \nonumber\\
                &&
                \;\;\;\;\;\;\;\;\;\;\;         + 2\, \Phi
                \partial^{\alpha\dot{\alpha}} \partial^{\beta\dot{\beta}}
                \Phi_{(\alpha\beta)(\dot{\alpha}\dot{\beta})}
                - 6 \Phi \Box \Phi \Big].\nonumber
                \eea
                It is invariant under the residual gauge freedom $\delta \Phi_{\beta \dot{\beta}\alpha \dot{\alpha} } =\frac{1}{2} \left(\partial_{\alpha\dot{\alpha}} a_{\beta\dot{\beta}}
                + \partial_{\beta\dot{\beta}} a_{\alpha\dot{\alpha}}  \right), \;\delta \Phi = \frac{1}{4} \partial_{\alpha\dot{\alpha}} a^{\alpha\dot{\alpha}}\,.$

\section{${\cal N}=2$ spin 3 }

        A generalization to higher integer spin ${\cal N}=2$ supermultiplets goes rather straightforwardly. The spin 3 example is significative.

            The spin 3 triad  of analytic gauge superfields is introduced as:
        \bea
                h^{++(\alpha\beta)(\dot\alpha\dot\beta)} (\zeta)\,, \;\;\; h^{++ \alpha\dot\alpha}(\zeta), \;\;\; h^{++(\alpha\beta)\dot{\alpha}+}(\zeta), \;\;\;
                h^{++(\dot\alpha\dot\beta){\alpha}+}(\zeta)\,, \nonumber
                \eea
            and has the following transformation laws, with the analytic gauge parameters:
            \bea
                && \delta h^{++(\alpha\beta)(\dot{\alpha}\dot{\beta})} = D^{++} \lambda^{(\alpha\beta)(\dot{\alpha}\dot{\beta})}
                + 2i \big[\lambda^{+(\alpha\beta)(\dot{\alpha}}  \bar{\theta}^{+\dot{\beta})}
                + \theta^{+(\alpha}  \bar{\lambda}^{+\beta)(\dot{\alpha}\dot{\beta})}\big],
                \nonumber \\
                && \delta h^{++\alpha\dot{\alpha}} = D^{++} \lambda^{\alpha\dot{\alpha}} - 2i  \big[\lambda^{+(\alpha\beta)\dot{\alpha}} \theta^{+}_{\beta} +
                \bar\lambda^{+(\dot\alpha\dot\beta){\alpha}} \bar\theta^{+}_{\dot\beta}\big], \nonumber \\
                && \delta h^{++(\alpha\beta)\dot{\alpha}+} = D^{++} \lambda^{+(\alpha\beta)\dot{\alpha}}\,, \quad\; \delta h^{++(\dot{\alpha}\dot{\beta})\alpha+}
                = D^{++} \lambda^{+(\dot{\alpha}\dot{\beta})\alpha}\,. \nonumber
                \eea

            The bosonic physical fields in  the Wess-Zumino gauge are collected in
            \bea
                h^{++(\alpha\beta)(\dot{\alpha}\dot{\beta})}
                &=&
                -2i \theta^{+\rho} \bar{\theta}^{+\dot{\rho}} \Phi^{(\alpha\beta)(\dot{\alpha}\dot{\beta})}_{\rho\dot{\rho}}% + \ldots \nonumber \\
                + \left[ (\bar{\theta}^+)^2 \theta^+ \psi^{(\alpha\beta)(\dot{\alpha}\dot{\beta})i}u^-_i + c.c.  \right] + \ldots \nonumber \\
                %&& + \,(\theta^+)^2 \bar{\theta}^+ \bar{\psi}^{(\alpha\beta)(\dot{\alpha}\dot{\beta})i}u_i^-
                %        +  (\theta^+)^4 V^{(\alpha\beta)(\dot{\alpha}\dot{\beta})(ij)}u^-_iu^-_j\,, \nonumber \\
                h^{++\alpha\dot{\alpha}} &=&
                -2i \theta^{+\rho} \bar{\theta}^{+\dot{\rho}} C^{\alpha\dot{\alpha}}_{\rho\dot{\rho}} + \ldots \nonumber
                %       + (\bar{\theta}^+)^2 \theta^+ \rho^{\alpha\dot{\alpha}i}u^-_i + (\theta^+)^2 \bar{\theta^+} \bar{\rho}^{\alpha\dot{\alpha}i}u_i^- \nonumber \\
                %        && + \,(\theta^+)^4 S^{\alpha\dot{\alpha}(ij)}u^-_iu^-_j\,, \nonumber \\
                %h^{++(\alpha\mu)\dot{\alpha}+} &=& (\theta^+)^2 \bar{\theta}^+_{\dot{\mu}} P^{(\alpha\mu)\dot{\alpha}\dot{\mu}}
                %       +  \left(\bar{\theta}^+\right)^2 \theta^+_\nu \left[\varepsilon^{\nu(\alpha} M^{\mu)\dot{\alpha}} + T^{\dot\alpha(\alpha\mu\nu)}\right] \nonumber \\
                %       && +\,  (\theta^+)^4 \chi^{(\alpha\mu)\dot{\alpha}i}u^-_i\,, \nonumber \\
                %        h^{++\alpha(\dot{\alpha}\dot{\mu})+} &=& \widetilde{\left(h^{++(\alpha\mu)\dot{\alpha}+}\right)}\,. \nonumber
                %\end{cases}
                \eea
             The physical gauge fields are $\Phi^{(\alpha\beta)(\dot{\alpha}\dot{\beta})}_{\rho\dot{\rho}}$ (spin 3 gauge field),
            $C^{\alpha\dot{\alpha}}_{\rho\dot{\rho}}$ (spin 2 gauge field) and $\psi^{(\alpha\beta)(\dot{\alpha}\dot{\beta})i}_\gamma$
            (spin 5/2 gauge field). The rest of fields are auxiliary. On shell, we end up with the $\mathcal{N}=2$ multiplet $( 3, 5/2, 5/2, 2)$.

        Some residual gauge freedom can be used to put the physical bosonic gauge fields into the irreducible form
            \bea
                && \Phi_{\gamma\dot\gamma (\alpha\beta)(\dot{\alpha}\dot{\beta})} = \Phi_{(\alpha\beta\gamma)(\dot\alpha\dot\beta\dot\gamma)}
                + \varepsilon_{\dot\gamma(\dot\alpha} \varepsilon_{\gamma(\beta} \Phi_{\alpha)\dot\beta)}\,, \nonumber \\
                &&  C_{\gamma\dot\gamma \alpha\dot\alpha} = C_{(\gamma\alpha)(\dot\gamma\dot\alpha)} + \varepsilon_{\gamma\alpha}\varepsilon_{\dot\gamma\dot\alpha} C\,, \nonumber
                \eea
            with the following gauge transformations
        \bea
                &&\delta \Phi_{(\alpha \gamma \beta ) (\dot{\alpha} \dot{\gamma} \dot{\beta} )}
                =
                \partial_{(\beta(\dot{\beta}} a_{\alpha \gamma) \dot{\alpha} \dot{\gamma})}\,, \quad \delta \Phi_{\alpha\dot{\beta}}
                = \frac{4}{9} \partial^{\gamma \dot{\gamma}} a_{(\alpha\gamma)(\dot{\beta}\dot{\gamma})}\,, \nonumber \\
                &&  \delta C_{(\alpha\beta) (\dot{\alpha}\dot{\beta})} = \partial_{(\beta (\dot{\beta}} b_{\alpha)\dot{\alpha})},
                \quad
                \delta C = \frac{1}{4} \partial_{\alpha\dot{\alpha}} b^{\alpha\dot{\alpha}}. \nonumber
                \eea
            %\vspace{0.5cm}
             These are just the correct gauge transformations for the Fronsdal spin 3 fields ($\Phi_{(\alpha\beta\gamma)(\dot\alpha\dot\beta\dot\gamma)}, \Phi_{\alpha\dot\beta}$)
            and spin 2 fields ($C_{(\alpha\beta) (\dot{\alpha}\dot{\beta})}, C $).

             The invariant superfield action is constructed literally on the pattern of the spin 2 case
            \bea
                S_{(s=3)} = \int d^4x d^8\theta du \,\Big\{G^{++(\alpha\beta)(\dot\alpha\dot\beta)}G^{--}_{(\alpha\beta)(\dot\alpha\dot\beta)}
                + G^{++\alpha\dot\beta}G^{--}_{\alpha\dot\beta} \Big\}, \nonumber
                \eea
            with
            \bea
                && G^{++(\alpha\beta)(\dot\alpha\dot\beta)} = h^{++(\alpha\beta)(\dot\alpha\dot\beta)}+ 2i\big[ h^{++(\alpha\beta)(\dot\alpha+} \bar\theta^{-\dot\beta)} -
                h^{++(\dot\alpha\dot\beta) (\alpha +}\theta^{-\beta)}\big]\,, \nonumber\\
                &&G^{++\alpha\dot\beta} = h^{++\alpha\dot\beta} - 2i \big[h^{++(\alpha\beta)\dot\beta+}\theta^-_\beta - \bar\theta^-_{\dot\alpha}h^{++(\dot\alpha\dot\beta)\alpha+} \big],
                \nonumber \\
                %&& \delta_\epsilon G^{\pm\pm(\alpha\beta)(\dot\alpha\dot\beta)}  = \delta_\epsilon G^{\pm\pm\alpha\dot\beta} =0\,, \lb{Scalsusy3} \\
                && D^{++} G^{--(\alpha\beta)(\dot\alpha\dot\beta)} - D^{--}G^{++(\alpha\beta)(\dot\alpha\dot\beta)} = 0\,, \; \nonumber \\
                && D^{++}G^{--\alpha\dot\beta} - D^{--}G^{++\alpha\dot\beta} = 0. \nonumber
                \eea

            The component spin 3 bosonic action reads
            \bea
                %\begin{split}
                % S_{(s= 3)} &= \int d^4x d^8\theta du \; \left[G^{++\alpha_1\alpha_2\dot{\alpha}_1\dot{\alpha}_2} G^{--\alpha_1\alpha_2\dot{\alpha}_1\dot{\alpha}_2} + 4 G^{++\alpha \dot{\alpha}} G^{--\alpha \dot{\alpha}} \right]
                % \\& =  -\int d^4x \; \frac{3}{2}\Phi^{(\alpha_1\alpha_2)\alpha_3 (\dot{\alpha}_1  \dot{\alpha}_2) \dot{\alpha}_3} \mathcal{G}_{(\alpha_1\alpha_2)\alpha_3 (\dot{\alpha}_1  \dot{\alpha}_2) \dot{\alpha}_3}
                %\\&=
                && S_{(s= 3)} =   \int d^4x\; \Big\{\Phi^{(\alpha_1\alpha_2\alpha_3)( \dot{\alpha}_1\dot{\alpha}_2\dot{\alpha}_3)}
                \Box \Phi_{(\alpha_1\alpha_2\alpha_3)( \dot{\alpha}_1\dot{\alpha}_2\dot{\alpha}_3)} \nonumber \\
                && \;\;\;\;\;\;\;\;\;\;\;\;\;\;\;\;\;- \,\frac{3}{2} \Phi^{(\alpha_1\alpha_2\alpha_3)( \dot{\alpha}_1\dot{\alpha}_2\dot{\alpha}_3)}
                \partial_{\alpha_1\dot{\alpha}_1} \partial^{\rho\dot{\rho}} \Phi_{(\rho\alpha_2\alpha_3)( \dot{\rho}\dot{\alpha}_2\dot{\alpha}_3)} \nonumber \\
                &&\;\;\;\;\;\;\;\;\;\;\;\;\;\;\;\;\;+ \,3 \Phi^{(\alpha_1\alpha_2\alpha_3)( \dot{\alpha}_1\dot{\alpha}_2\dot{\alpha}_3)} \partial_{\alpha_1\dot{\alpha}_1} \partial_{\alpha_2\dot{\alpha}_2}
                \Phi_{\alpha_3\dot{\alpha}_3}
                - \frac{15}{4} \Phi^{\alpha\dot{\alpha}} \Box  \Phi_{\alpha\dot{\alpha}} \nonumber \\
                && \;\;\;\;\;\;\;\;\;\;\;\;\;\;\;\;\;+\,
                \frac{3}{8} \partial_{\alpha_1\dot{\alpha}_1} \Phi^{\alpha_1\dot{\alpha}_1} \partial_{\alpha_2\dot{\alpha}_2} \Phi^{\alpha_2\dot{\alpha}_2}\Big\}. \nonumber
                %\end{split}
                \eea

\section{${\cal N}=2$ spin s }

            The general case with the maximal spin $ s$ is spanned by the following analytic gauge potentials
        \bea
                h^{++\alpha(s-1)\dot\alpha(s-1)}(\zeta),\;\;\; h^{++\alpha(s-2)\dot\alpha(s-2)}(\zeta),\;\;\; h^{++\alpha(s-1)\dot\alpha(s-2)+}(\zeta),\;\;\;
                h^{++\dot\alpha(s-1)\alpha(s-2)+}(\zeta), %\nonumber
                \eea
            where $\alpha(s) := (\alpha_1 \ldots \alpha_s), \dot\alpha(s) := (\dot\alpha_1 \ldots \dot\alpha_s)$.

            The relevant gauge transformations can also be defined and shown to leave, in the WZ-like gauge, the  physical field multiplet
            $({ s, s-1/2, s-1/2, s-1})$.

             The invariant action has the universal form for any ${\bf s}$
            \bea
                && S_{(s)} = (-1)^{s+1} \int d^4x
                d^8\theta du \,\Big\{G^{++\alpha(s-1)\dot\alpha(s-1)}G^{--}_{\alpha(s-1)\dot\alpha(s-1)} \nonumber \\
                &&\;\;\;\;\;  \,\qquad\qquad\qquad\qquad\qquad\qquad +\,
                G^{++\alpha(s-2)\dot\alpha(s-2)}G^{--}_{\alpha(s-2)\dot\alpha(s-2)}
                \Big\},
                \eea
            where
        \bea
                && G^{\pm\pm\alpha(s-1)\dot\alpha(s-1)} = h^{\pm\pm\alpha(s-1)\dot\alpha(s-1)} + 2i \big[h^{\pm\pm\alpha(s-1)(\dot\alpha(s-2)+}\bar\theta^{-\dot\alpha_{s-1})} \nonumber \\
                && \;\;\;\;\;\;\;\;\;\;\;\;\;\;\;\;\;\;\;\;\;\;\;\;\;\;\qquad- h^{\pm\pm\dot\alpha(s-1)(\alpha(s-2)+}\,\theta^{-\alpha_{s-1})} \big], \nonumber \\
                && G^{\pm\pm\alpha(s-2)\dot\alpha(s-2)} = h^{\pm\pm\alpha(s-2)\dot\alpha(s-2)} - 2i \big[h^{\pm\pm(\alpha(s-2)\alpha_{s-1}) \dot\alpha(s-2)+}\theta^{-}_{\alpha_{s-1}} \nonumber \\
                && \;\;\;\;\;\;\;\;\;\;\;\;\;\;\;\;\;\;\;\;\;\;\;\;\;\;\qquad+ h^{\pm\pm\alpha(s-2)(\dot\alpha(s-2)\dot\alpha_{(s-1)})+}\,\bar\theta^{-}_{\alpha_{s-1}} \big], \nonumber
                \eea
            and the negatively charged potentials are related to the basic ones by the appropriate harmonic zero-curvature conditions.

\section{Hypermultiplet couplings}

         The construction of interactions in the theory of
            higher spins is a very important task \cite{Bekaert:2004qos, Didenko:2014dwa}. The simplest higher spin interaction is described by a cubic vertex,
            e.g., bilinear in the matter fields and of the first order
            in gauge fields. At present, there is an extensive literature
            related to the construction of cubic higher spin interactions
            (e.g.,  \cite{Bengtsson:1983pd, Metsaev:2007rn, Manvelyan:2009vy, Manvelyan:2010wp} and many others).

         Supersymmetric $\mathcal{N}=1$ generalizations of the purely bosonic
            cubic vertices with matter and the corresponding supercurrents were
            explored in terms of $\mathcal{N}=1$ superfields in the papers \cite{Buchbinder:2017nuc,
                Buchbinder:2018wzq}.

             In \cite{Buchbinder:2022kzl} we have constructed, for the first time, the off-shell
            manifestly $\mathcal{N}=2$ supersymmetric cubic couplings
            $\mathbf{\frac{1}{2}-\frac{1}{2} - s}$  of an arbitrary higher
            integer  superspin $\mathbf{s}$ gauge $\mathcal{N}=2$ multiplet to the
            hypermultiplet matter in $4D, \mathcal{N}=2$ harmonic
            superspace.

         Our starting point is the ${\cal N}=2$ hypermultiplet off-shell free action:

                \begin{equation}
                    S = \int  d\zeta^{(-4)}  \; \mathcal{L}^{+4}_{free} = -\int d\zeta^{(-4)}  \; \frac{1}{2} q^{+a} \mathcal{D}^{++} q^+_a,\;\;\;\quad a = 1,2.\nonumber
            \end{equation}
        Here there is addition central charge term in harmonic derivative:
        \begin{equation*}
            \mathcal{D}^{++}  = D^{++} + \left((\theta^+)^2 - (\bar{\theta}^+)^2\right) \partial_5\,.
        \end{equation*}
        This term is used for description of massive hypermultiplet.

        \subsection{Spin 1 coupling}

             We reproduce the gauge higher-spin ${\cal N}=2$ superfields from gauging the appropriate higher-derivative rigid (super)symmetries of this free
            hypermultiplet action. The simplest symmetry is of zero order in derivatives, it is $U(1)$ transformation of $q^{+a}$,

                \begin{equation}
                    \delta q^{+a} = -\lambda_0 J q^{+ a}, \quad J q^{+ a} = i (\tau_3)^a_{\;b} q^{+b}\,.\nonumber
            \end{equation}

            Gauging of this symmetry is just replacing $\lambda_0$ by analytic superparameter, $\lambda_0 \rightarrow  \lambda(\zeta)$, and in order to make the
            $q^{+a}$ action gauge-invariant, we perform the change
                \bea
                 {\cal D}^{++} \;\Rightarrow \;\mathfrak{D}^{++}_{(1)} =  {\cal D}^{++} + h^{++}J\,,
                 \qquad
                \delta_\lambda h^{++} = {\cal D}^{++}\lambda\,. \nonumber
                \eea
            So the $q^+$ gauge invariant action is:
                \begin{equation}
                \mathcal{L}^{+4}_{free}\;\;\to\;\; \mathcal{L}^{+4 (s=1)}_{gauge} = - \frac{1}{2} q^{+a} \left(\mathcal{D}^{++} + h^{++} J \right) q^+_a\,.%\nonumber
            \end{equation}

            There is no direct relation between $J$ and $\partial_5$: one can choose $\partial_5 q^{+a}=0$ which corresponds to
            massless $q^{+a}$
            or $\partial_5 q^{+a} \sim   mJ q^{+ a}$, which
            corresponds to massive $q^{+a}$.

\subsection{Spin 2 coupling}

 The global symmetry to be gauged in the spin 2 case looks somewhat more complicated, it is of first order in derivatives

                \begin{equation}
                    \delta_{rig} q^{+a} =  -\hat{\Lambda}_{rig} q^{+a},\nonumber
            \end{equation}\begin{eqnarray}
                    \hat{\Lambda}_{rig} &=& \left( \lambda^{\alpha\dot{\alpha}} - 2i \lambda^{-\alpha} \bar{\theta}^{+\dot{\alpha}} - 2i \theta^{+\alpha} \bar{\lambda}^{-\dot{\alpha}}  \right) \partial_{\alpha\dot{\alpha}}
                    + \lambda^{+\alpha} \partial^-_{\alpha} + \bar{\lambda}^{+\dot{\alpha}} \partial^-_{\dot{\alpha}} \nonumber \\
                    &&+\, \left( \lambda^5 + 2i \lambda^{\hat{\alpha}-} \theta^+_{\hat{\alpha}} \right) \partial_5  := \Lambda^M \partial_M\,, \quad [{\cal D}^{++},\hat{\Lambda}_{rig}] = 0\,.\nonumber
            \end{eqnarray}
            It involves five constant bosonic parameters $\lambda^{\alpha\dot{\alpha}}$, $\lambda^5$, four constant spinor parameters
            $\lambda^{\pm\hat{\alpha}} = \lambda^{\hat{\alpha}i} u^\pm_i$
            and can be treated as a copy of the rigid ${\cal N}=2$ supersymmetry transformations in their active form.
            We gauge just these transformations, leaving ${\cal N}=2$ supersymmetry realized on the coordinates rigid.

            In this case there are two possibilities for gauge transformations of the hypermultiplet:
            \begin{equation}
                    \delta_1 q^{+a} =  -\hat{\Lambda}_{(2)} q^{+a},
                    \;\;\;\;\;
                    \hat{\Lambda}_{(2)}: = \lambda^M \partial_M = \lambda^{\alpha\dot{\alpha}} \partial_{\alpha\dot{\alpha}}
                    + \lambda^{+\alpha} \partial^-_{\alpha} + \bar{\lambda}^{+\dot{\alpha}} \partial^-_{\dot{\alpha}} + \lambda^5 \partial_5\,,\nonumber
            \end{equation}
                \begin{equation}
                    \delta_2 q^{+a} =  -\frac{1}{2} \Omega_{(2)} q^{+a},
                    \;\;\;\;\;
                    \Omega_{(2)} := (-1)^{P(M)} \partial_M \lambda^M = \partial_{\alpha\dot{\alpha}}\lambda^{\alpha\dot{\alpha}}
                    - \partial^-_{\alpha} \lambda^{+\alpha}  - \partial^-_{\dot{\alpha}}\bar{\lambda}^{+\dot{\alpha}}.\nonumber
            \end{equation}
                \begin{equation}
                    \left(\delta_1 + \delta_2 \right) \mathcal{L}^{+4}_{free}
                    =
                    \frac{1}{2} q^{+a} [\mathcal{D}^{++}, \hat{\Lambda}_{(2)}] q^+_a. \nonumber
            \end{equation}

            Next, we introduce the $\mathcal{N}=2$ invariant differential operator
            \begin{equation}
                    \hat{\mathcal{H}}^{++}_{(2)} = h^{++\alpha\dot{\alpha}} \partial_{\alpha\dot{\alpha}} + h^{++\hat{\mu}+} \partial^-_{\hat{\mu}} + h^{++5}\partial_5\,,\nonumber
            \end{equation}
            with the gauge transformation law
                \begin{equation}
                    \delta \hat{\mathcal{H}}^{++}_{(2)} = [\mathcal{D}^{++}, \hat{\Lambda}_{(2)}].\nonumber
            \end{equation}
            Then the linear in gauge superfields part of the gauge-invariant action reads
            \begin{equation}
                    \mathcal{L}^{+4}_{free}\;\;\to\;\; \mathcal{L}^{+4 (s=2)}_{gauge} = - \frac{1}{2} q^{+a} \left(\mathcal{D}^{++} + \hat{\mathcal{H}}^{++}_{(2)} \right) q^+_a\,.%\nonumber
            \end{equation}

            In fact $\mathcal{L}^{+4 (s=2)}_{gauge}$ can be made fully gauge invariant (and not at the linearized level) by deforming the gauge
            transformation law of $\hat{\mathcal{H}}^{++}_{(2)}$ as
        \begin{equation}
                    \delta_{full} \hat{\mathcal{H}}^{++}_{(2)}  =
                    [\mathcal{D}^{++} +\hat{\mathcal{H}}^{++}_{(2)}, \hat{\Lambda}_{(2)}]\,.\nonumber
            \end{equation}

             A complete nonlinear harmonic superfield action of ${\cal N}=2$ supergravity including the pure supergravity part was constructed in \cite{Galperin:1987em}.

             \subsection{Spin 3 coupling}

     A surprising peculiarity of the superspin 3 case is that the relevant rigid two-derivative transformations to be gauged and
            the resulting couplings of the relevant gauge superfields to the hypermultiplet can be consistently defined only at cost of breaking rigid $SU(2)_{PG}$ symmetry
            down to $U(1)$ generator which is manifestly present in all formulas. This peculiarity extends to all odd ${\cal N}=2$ spins.

         Though in this case from the beginning one can define 4 independent transformations of $q^{+ a}$, only two their fixed combinations can be compensated
            by the appropriate transformations of gauge superfields:
            \begin{eqnarray}
                    \delta_{\lambda} q^{+a} &=& -\left[ \lambda^{\alpha\dot\alpha M}\partial_M\partial_{\alpha\dot\alpha} + \frac12 ( \partial_{\alpha\dot\alpha}\lambda^{\alpha\dot\alpha M})\partial_M
                    +
                    \frac12 \Omega^{\alpha\dot{\alpha}} \partial_{\alpha\dot{\alpha}}
                    + \frac12\Omega_{(3)}\right] (\tau_3)^a_{\;b} q^{+ b}\,,\nonumber\\
                    && \delta_{\xi} q^{+a} =  - \xi \,\Omega_{(3)}\,J q^{+a} = - i\xi \,\Omega_{(3)}\,(\tau_3)^a_{\;b} q^{+ b}\,,\nonumber
            \end{eqnarray}
            where
            \begin{eqnarray}
                    && \lambda^{\alpha\dot\alpha M}\partial_M = \lambda^{(\alpha\beta)(\dot{\alpha}\dot{\beta})}\partial_{\beta\dot{\beta}} + \lambda^{(\alpha\beta)\dot{\alpha}+} \partial^-_\beta +
                    \bar{\lambda}^{(\dot{\alpha}\dot{\beta})\alpha+} \partial^-_{\dot{\beta}} + \lambda^{\alpha\dot{\alpha}} \partial_5\,, \nonumber\\
                    &&\Omega_{(3)} = (\partial_{\alpha\dot\alpha}\Omega_{(3)}^{\alpha\dot\alpha})\,, \quad \Omega^{\alpha\dot\alpha} = (-1)^{P(M)}(\partial_M\lambda^{\alpha\dot\alpha M})\,. \nonumber
            \end{eqnarray}

            Defining
            \begin{eqnarray}
                    && \hat{\mathcal{H}}^{++\alpha\dot{\alpha}} = h^{++(\alpha\beta)(\dot{\alpha}\dot{\beta})}\partial_{\beta\dot{\beta}} + h^{++(\alpha\beta)\dot{\alpha}+} \partial^-_\beta
                    +
                    \bar{h}^{++(\dot{\alpha}\dot{\beta})\alpha+} \partial^-_{\dot{\beta}} + h^{++\alpha\dot{\alpha}} \partial_5\,, \nonumber \\
                    && \Gamma^{++\alpha\dot{\alpha}} = \partial_{\beta\dot{\beta}}h^{++(\alpha\beta)(\dot{\alpha}\dot{\beta})}-
                    \partial^-_{\beta}h^{++(\alpha\beta)\dot{\alpha}+} -
                    \partial^-_{\dot{\beta}}
                    h^{++\alpha(\dot{\alpha}\dot{\beta})+}, \nonumber \\
                    &&  \hat{\mathcal{H}}^{++}_{(3)}
                    = \hat{\mathcal{H}}^{++\alpha\dot{\alpha}} \partial_{\alpha\dot{\alpha}}, \qquad\; \Gamma^{++}_{(3)} = \partial_{\alpha\dot{\alpha}}\Gamma^{++\alpha\dot{\alpha}}\, \nonumber \\
                    &&\delta \hat{\mathcal{H}}^{++} = [\mathcal{D}^{++}, \hat{\Lambda}_{(3)}]\,, \quad\,
                    \hat{\Lambda}_{(3)} = {\lambda}^{\alpha\dot{\alpha}M}\partial_M \partial_{\alpha\dot{\alpha}}\,, \quad\;
                    \delta\Gamma^{++}_{(3)} = {\cal D}^{++}\Omega_{(3)}\,,  \nonumber
            \end{eqnarray}
            we obtain a gauge invariant extension of the $q^+$ action as
            \begin{eqnarray}
                        \mathcal{L}^{+4}_{free}\;\;\to\;\; \mathcal{L}^{+4(s=3)}_{gauge} =
                    -\frac12q^{+a} \left(\mathcal{D}^{++} + \hat{\mathcal{H}}^{++}_{(3)} J +\xi \Gamma^{++}_{(3)} J  \right) q^+_a\,.%\nonumber
            \end{eqnarray}
            The presence of constant $\xi$ in the gauged Lagrangian  shows that off shell there are 2 types of possible interactions of the ${\cal N}=2$ spin
            ${\bf 3}$
            with the hypermultiplet. The coefficient $\xi$ is a dimensionless coupling constant that measures the relative strength of these interactions. Actually, we have checked that
            on shell the $\xi$ term does not contribute to cubic vertex (at least in the bosonic sector), it survives only off-shell and perhaps can play some role in the quantum theory.

\subsection{Spin 4 coupling}

             In the superspin \textbf{4} case the transformations of $q^{+a}$ do not require internal symmetry generators, so they preserve $SU(2)_{PG}$ invariance.
            The rigid symmetry
            transformations are of third order in derivatives and well defined. Their localization in the analytic subspace admit 6 independent variations, but  only 3 of them
            turn out to finally  matter
            \begin{eqnarray}
                    && \delta_1 q^{+a} = -\partial_{\alpha\dot{\alpha}} \partial_{\beta\dot{\beta}} \hat{\Lambda}^{\alpha\beta\dot{\alpha}\dot{\beta}} q^{+a}\,,\quad \;
                    \delta_2 q^{+a} =  -\hat{\Lambda}^{\alpha\beta\dot{\alpha}\dot{\beta}} \partial_{\alpha\dot{\alpha}} \partial_{\beta\dot{\beta}} q^{+a}\,,\quad  \nonumber %\\
                %   &&
                    \delta_3 q^{+a} = -\partial_{\alpha\dot{\alpha}} \partial_{\beta\dot{\beta}} \Omega^{\alpha\beta\dot{\alpha}\dot{\beta}} q^{+a}\,, \nonumber
            \end{eqnarray}
            where
            \begin{eqnarray}
                    \hat{\Lambda}^{\alpha\beta\dot{\alpha}\dot{\beta}} = \lambda^{(\alpha\beta)(\dot{\alpha}\dot{\beta})M}\partial_M\,, \quad\;
                    \Omega^{\alpha\beta\dot{\alpha}\dot{\beta}} = (-1)^{P(M)}(\partial_M\lambda^{(\alpha\beta)(\dot{\alpha}\dot{\beta})M}) \nonumber
            \end{eqnarray}
            and derivatives freely act to the right.
             One can calculate
            \begin{equation}
                    \left(\delta_1 + \delta_2  + \delta_3\right) \mathcal{L}^{+4}_{free}
                    =
                    \frac{1}{2} q^{+a} [\mathcal{D}^{++}, \hat{\Lambda}_{(4)}] q^+_a. \nonumber
            \end{equation}

         Then the modified $q^{+a}$ action reads
            %gauge invariant up to 1st order in gauge superfields reads
        \begin{equation}
                    %\begin{split}
                    \mathcal{L}^{+4}_{free}\;\;\to\;\;  \mathcal{L}^{+4(s=4)}_{gauge} = -   \frac12\, q^{+a}  \left(\mathcal{D}^{++} + \hat{\mathcal{H}}^{++}_{(4)} \right) q^+_a\,,%\nonumber
            \end{equation}
            with
        \begin{eqnarray}
                     \hat{\mathcal{H}}^{++}_{(4)} = h^{++(\alpha\beta)(\dot{\alpha}\dot{\beta})M}\partial_M\partial_{\alpha\dot{\alpha}}\partial_{\beta\dot{\beta}}\,,
                     \quad
                     \delta \hat{\mathcal{H}}^{++}_{(4)} = [{\cal D}^{++},\hat{\Lambda}_{(4)}]\,,\quad \hat{\Lambda}_{(4)}=
                    \lambda^{(\alpha\beta)(\dot{\alpha}\dot{\beta})M}\partial_M\partial_{\alpha\dot{\alpha}}\partial_{\beta\dot{\beta}}\,.\nonumber
                    \eea

    \subsection{Generaliations}

             The hypermultiplet couplings of the ${\cal N}=2$ gauge multiplets of higher superspins ${\bf s}$ can be constructed quite analogously and have the uniform structure:
                gauge superfields and gauge parameters are the appropriate differential
                operators of the rank ${\bf s -1}$.

                 Everything can be easily extended to a system of several hypermultiplets.
                The free Lagrangian of $n$ hypermultiplets can be written in the manifestly ${\rm USp}(2n)$ invariant form as
                \bea
                    {\cal L}^{+4}_{free, n} = \frac12 q^{+ A}{\cal D}^{++}q^+_A\,, \qquad \widetilde{q^+_A} = \Omega^{AB} q^+_B\,, \quad\;\; A = 1, 2, \ldots , 2n\,,\nonumber
                    \eea
                where $\Omega^{AB} = -\Omega^{BA}$ is ${\rm USp}(2n)$ invariant constant $2n \times 2n$ symplectic metric. It can be rewritten in an equivalent
                complex form as\bea
                    {\cal L}^{+4}_{free, n} \sim \tilde{q}^{+a}{\cal D}^{++}q^+_a -{\cal D}^{++} \tilde{q}^{+a} q^+_a\,, \quad  a= 1, 2, \ldots , n\,, \;\;q^+_A = (q^+_a, - \tilde{q}^{+ a})\, \nonumber
                    \eea
                so that the manifest symmetry is ${\rm U}(n)= {\rm SU}(n)\times {\rm U}(1) \subset {\rm USp}(2n)$, with respect to which $q^+_a$ and $\tilde{q}^{+ a}$
                transform in the fundamental and co-fundamental representations. Then one can identify the $U(1)$ generator needed for description of odd spins as
            \bea
                    J q^+_a  = i q^+_a\,, \quad J \tilde{q}^{+ a} =  -i \tilde{q}^{+ a}\,, \nonumber
                    \eea
                so that ${\rm USp}(2n)$ gets broken to ${\rm U}(n)$. Some other options with ${\rm SU}(n)$ also broken, are as well admissible.

\section{Summary and outlook}
        The theory of ${\cal N}= 2$ higher spins  $s\geq 3$ opens a new promising direction
        of applications of the harmonic superspace approach which earlier proved to be indispensable for description of
        more conventional ${\cal N}= 2$ theories with the maximal spins $s\leq 2$. Once again,
        the basic property underlying these new higher-spin theories is the harmonic Grassmann analyticity (all basic gauge potentials
        are unconstrained analytic superfields involving an infinite number of degrees of freedom off shell before fixing WZ-type gauges).
        \vspace{0.4cm}

        {\bf Some further lines of development:}:

        \begin{itemize}
            \item The natural next steps are the construction and investigation of $4D, \mathcal{N}=2$ higher-spin supercurrents of the hypermultiplet and more detailed
            study of the component structure;

            \item Construction of ${\cal N}=2$ supersymmetric half-integer spin
            theories;

            \item An extension to AdS  background;

            \item From the linearized theory to its full nonlinear version? At present, the latter is known only for $s\leq 2$ (${\cal N}=2$ super Yang - Mills
            and ${\cal N}=2$ supergravities). This problem will seemingly require  accounting  for all higher ${\cal N}=2$ superspins simultaneously. New supergeometries?

    \end{itemize}

\textbf{Acknowledgments}

\medskip

Work of I.B. and E.I.
	was supported in part by the Ministry of Education of the Russian
	Federation, project FEWF-2020-0003.
	Work of N.Z. was partially supported by by the grant 22-1-1-42-2 from the Foundation for the Advancement of Theoretical Physics and Mathematics “BASIS”.

\end{document}